\documentclass[12pt]{iopart}
\usepackage{epsfig} 
\usepackage{amssymb}
\begin{document}

\title{Status report of the baseline collimation system of CLIC. Part~I}

\author{J.~Resta-L\'opez$^1$, D.~Angal-Kalinin$^2$, B.~Dalena$^3$, J.~L.~Fern\'andez-Hernando$^2$, F.~Jackson$^2$, D.~Schulte$^3$, A.~Seryi$^1$ and R.~Tom\'as$^3$}

\address{$^1$JAI, University of Oxford, UK}
\address{$^2$STFC, Daresbury, UK}
\address{$^3$CERN, Geneva, Switzerland}

\ead{j.restalopez@physics.ox.ac.uk}

\begin{abstract}
Important efforts have recently been dedicated to the characterisation and improvement of the design of the post-linac collimation system of the Compact Linear Collider (CLIC). This system consists of two sections: one dedicated to the collimation of off-energy particles and another one for betatron collimation. The energy collimation system is further conceived as protection system against damage by errant beams. In this respect, special attention is paid to the optimisation of the energy collimator design. The material and the physical parameters of the energy collimators are selected to withstand the impact of an entire bunch train. Concerning the betatron collimation section, different aspects of the design have been optimised: the transverse collimation depths have been recalculated in order to reduce the collimator wakefield effects while maintaining a good efficiency in cleaning the undesired beam halo; the geometric design of the spoilers has been reviewed to minimise wakefields; in addition, the optics design has been optimised to improve the collimation efficiency. This report presents the current status of the the post-linac collimation system of CLIC. Part~I of this report is dedicated to the study of the CLIC energy collimation system.  
\end{abstract}

\maketitle
\section{Introduction}
\label{intro}

The post-linac collimation systems of the future linear colliders will play an essential role in reducing the detector background at the interaction point (IP), and protecting the machine by minimising  the activation and damage of sensitive accelerator components. 

\begin{figure}[htb]
\begin{center}
\includegraphics*[width=14cm]{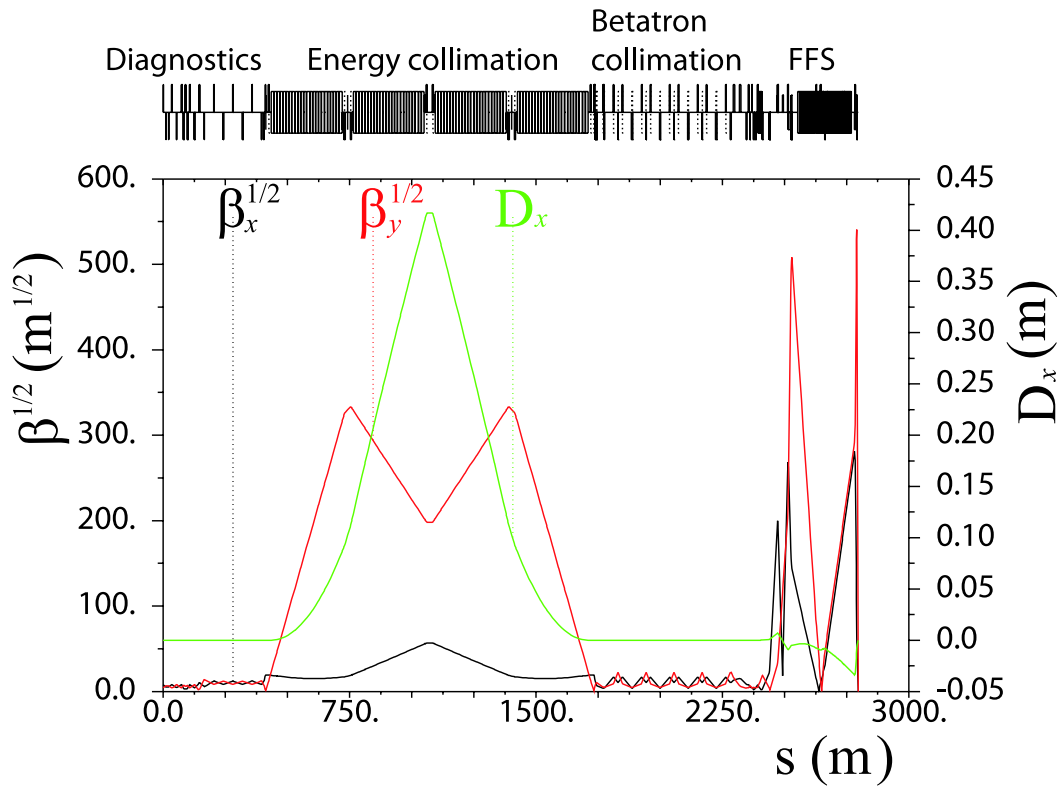}
\caption{Optical functions of the CLIC beam delivery system.}
\label{latticecoll}
\end{center}
\end{figure}

The CLIC Beam Delivery System (BDS), downstream of the main linac, consists of a 370 m long diagnostics section, an almost 2000 m long collimation system, and a 460 m long Final Focus System (FFS) \cite{Rogelio, Tecker}. Figure~\ref{latticecoll} shows the betatron and dispersion functions along the CLIC BDS. Some relevant CLIC design parameters are shown in Table~\ref{CLICparametros} for the options at 500 GeV and 3 TeV centre-of-mass (CM) energy.

\begin{table}[!htb]
\caption{CLIC parameters at 0.5~TeV and 3~TeV CM energy.}
\label{CLICparametros}
\begin{center}
\begin{tabular}{lcc}
\hline
Parameter & {\bf CLIC 0.5~TeV} & {\bf CLIC 3~TeV} \\
\hline \hline
Design luminosity ($10^{34}$ cm$^{-2}$s$^{-1}$) & 2.3 & 5.9 \\
Linac repetition rate (Hz) & 50 & 50  \\
Particles/bunch at IP ($\times 10^{9}$) & 6.8 & 3.72 \\
Bunches/pulse & 354 & 312 \\
Bunch length ($\mu$m) & 72 & 44 \\
Bunch separation (ns) & 0.5 & 0.5 \\
Bunch train length (ns) & 177 & 156 \\
Emittances $\gamma \epsilon_{x}$/$\gamma \epsilon_{y}$ (nm rad) & 2400/25 & 660/20 \\
Transverse beam sizes at IP $\sigma^{*}_x$/$\sigma^{*}_y$ (nm) & 202/2.3 & 45/0.9 \\
BDS length (km) & 1.73 & 2.79 \\
\hline
\end{tabular}
\end{center}
\end{table}

In the CLIC BDS there are two collimation sections:

\begin{itemize}
\item The first post-linac collimation section is dedicated to energy collimation. The energy collimation depth is determined by failure modes in the linac \cite{DanielandFZ}. A spoiler-absorber scheme (Fig.~\ref{spoilerabsorberscheme}), located in a region with non-zero horizontal dispersion, is used for intercepting miss-steered or errant beams with energy deviation larger than $1.3\%$ of the nominal beam energy.    

\item Downstream of the energy collimation section, a dispersion-free section, containing eight spoilers and eight absorbers, is dedicated to the cleaning of the transverse halo of the beam, thereby reducing the experimental background at the IP. 
\end{itemize}

\noindent The spoilers are thin devices ($\lesssim 1$ radiation length) which scrape the beam halo and, if accidentally struck by the full power beam, will increase the volume of the phase space occupied by the incident beam via multiple Coulomb scattering. In this way, the transverse density of the scattered beam is reduced for passive protection of the downstream absorber. The absorbers are usually thick blocks of material (of about 20 radiation length) designed to provide efficient halo absorption or complete removal of potentially dangerous beams.    

The optics of the CLIC collimation system was originally designed by rescaling  of the optics of the collimation system of the previous Next Linear Collider (NLC) project at 1 TeV centre-of-mass energy \cite{PT2, Assmann1} to the 3~TeV CLIC requirements. In the present CLIC baseline optics the length of the energy collimation section has been scaled by a factor 5 and the bending angles by a factor $1/12$ with respect to the 1 TeV NLC design \cite{Assmann2}. On the other hand, the optics of the CLIC betatron collimation section was not modified with respect to the original design of the NLC. 

It is worth mentioning that, unlike the International Linear collider (ILC) \cite{ILC}, where the betatron collimation section is followed by the energy collimators, in CLIC the energy collimation section is upstream of the betatron one. The main reason of choosing this lattice structure is because miss-phased or unstable off-energy drive beams are likely failure modes in CLIC, and they are expected to be much more frequent than large betatron oscillations with small emittance beams. Therefore, the energy collimation system is conceived as the first post-linac line of defence for passive protection against off-energy beams in the CLIC BDS.

Recently many aspects of the CLIC collimation system design have been reviewed and optimised towards a consistent and robust system for the Conceptual Design Report of CLIC (CLIC CDR), to be completed during 2011. In this report we describe the current status of the CLIC collimation system at 3~TeV CM energy. Here we mainly focus on the description of the collimation layout and the optimisation of the necessary parameters of the baseline design to improve the collimation performance, only taking into account the primary beam halo. The aim is to define basic specifications of the design. Studies including secondary particle production and muon collimation are described elsewhere \cite{Agapov, Deacon}. Part~I of this report is dedicated to the study of the CLIC energy collimation system.         
\section{Energy collimation}
\label{energysection}

The beam power of the CLIC beam in the BDS with nominal parameters at 3~TeV CM energy is about 14~MW. The sustained disposal of such a high beam power during beam operation is a challenging task. Operation failures might generate errant beams which can hit and damage machine components. Therefore, machine protection, based on active and passive strategies, is required. The general CLIC Machine protection strategies are described in \cite{JonkerIPAC10}. 

The CLIC energy collimation section is conceived to fulfil a function of passive protection in the BDS against miss-steered or errant beams coming from the main linac. The energy collimation depth is determined by fast failure modes which result in a significant energy deviation of the beam. For instance, possible CLIC fast (`in flight') failure modes scenarios can be caused by the effect of a missing drive beam, injection phase errors and changes in the charge of the main beam \cite{DanielandFZ}.  

\begin{figure}[htb]
\begin{center}
\includegraphics*[width=14cm]{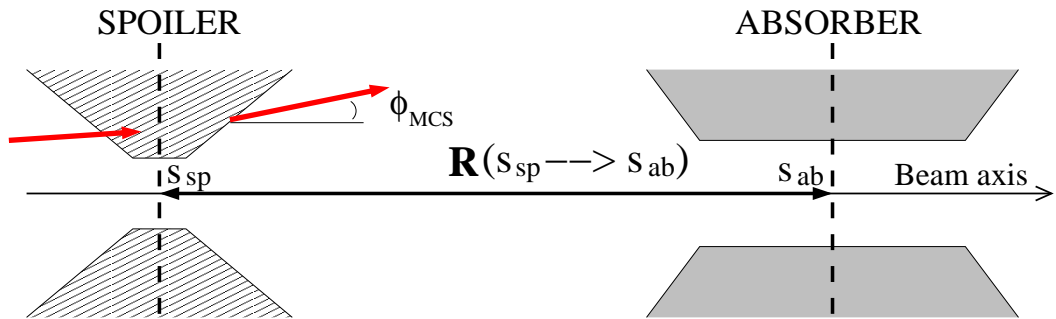}
\caption{Basic spoiler-absorber scheme.}
\label{spoilerabsorberscheme}
\end{center}
\end{figure}

\begin{figure}[htb]
\begin{center}
\includegraphics*[width=14cm]{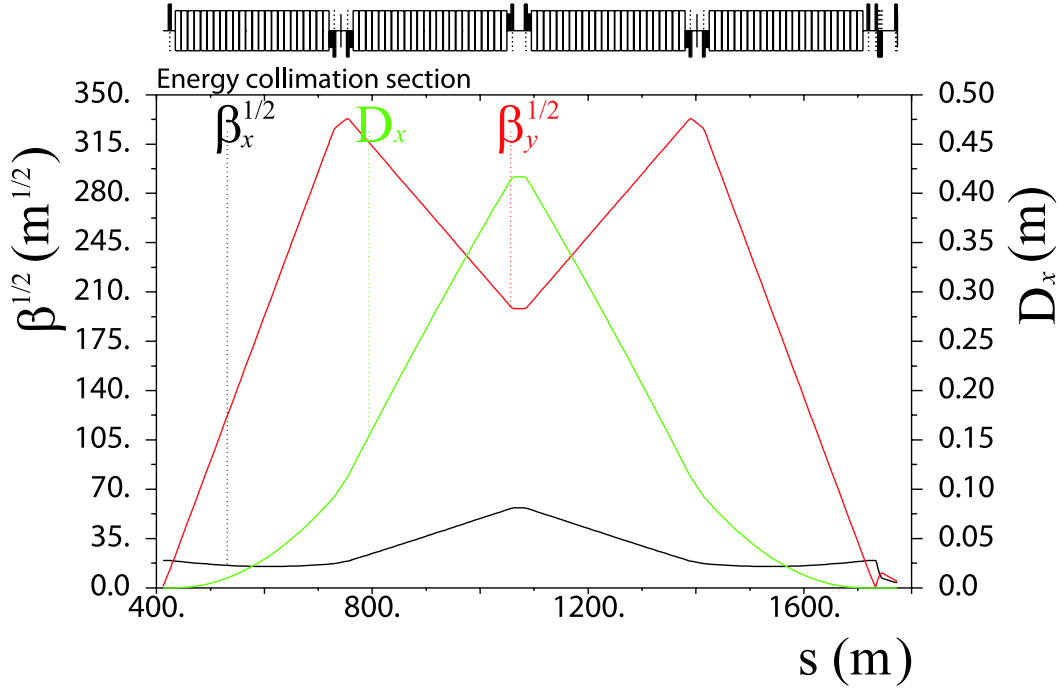}
\caption{Optical functions of the CLIC energy collimation section: horizontal dispersion and square root of the betatron functions.}
\label{Ecolloptics}
\end{center}
\end{figure}

The CLIC energy collimation system consists of a spoiler-absorber scheme (see Fig.~\ref{spoilerabsorberscheme}), located in a region with non-zero horizontal dispersion. The lattice layout of the CLIC energy collimation section is shown in Fig.~\ref{Ecolloptics}. The corresponding optical parameters and transverse beam size at the energy spoiler and absorber are indicated in Table~\ref{tablecoll3}. 

\begin{table}[!htb]
\caption{Optics and beam parameters at collimator position for energy collimation: longitudinal position ($s$), horizontal and vertical $\beta$-functions ($\beta_x$ and $\beta_y$), horizontal dispersion ($D_x$), horizontal and vertical rms beam sizes ($\sigma_x$ and $\sigma_y$). In this case a uniform energy distribution with 1$\%$ full width energy spread has been considered.}
\label{tablecoll3}
\begin{center}
\begin{tabular}{|l|c|c|c|c|c|c|}
\hline 
Name & $s$ [m] & $\beta_{x}$ [m] & $\beta_{y}$ [m] & $D_x$ [m] & $\sigma_x$ [$\mu$m] & $\sigma_y$ [$\mu$m] \\
\hline \hline
ENGYSP (spoiler) & 907.098 & 1406.33 & 70681.87 & 0.27 & 779.626 & 21.945   \\
ENGYAB (absorber) & 1072.098 & 3213.03 & 39271.54 & 0.416 &  1201.189 & 16.358  \\
\hline 
\end{tabular}
\end{center}
\end{table}

The selection of the material to make the spoiler is basically determined by the electrical, thermal and mechanical properties of the material. Regarding the survival condition of the energy spoiler, the robustness of the material is crucial. At the same time, a spoiler with high electrical conductivity is desired to avoid intolerable wakefield effects. Earlier studies of the CLIC spoiler heating and spoiler damage limit \cite{Fartouk} concluded that a spoiler made of beryllium (Be) might be a suitable solution in terms of high robustness and acceptable wakefields. On the other hand, in the current design the CLIC absorbers are made of titanium alloy ($90\%$ Ti, $6\%$ Al, $4\%$ V) with copper (Cu) coating.  

The collimation depth of the spoiler has been set to intercept beams with energy deviation larger than 1.3$\%$ of the nominal beam energy. The horizontal aperture for the energy collimator is then set to $a_x=D_x \delta_{\rm aper}$, with $D_x$ the horizontal dispersion at the spoiler position and $\delta_{\rm aper}=\pm 1.3\%$. 

It is necessary to point out that the energy collimation system, with a total length of 1400 m, is the longest part of the BDS. This space is filled almost entirely with bending magnets to generate the required horizontal dispersion. The length of the energy collimation system is determined by a trade-off between the following requirements: 

\begin{itemize}

\item The beam spot size at the collimators must be sufficiently large for passive protection. The energy collimators are required to withstand the impact of a full bunch train of nominal emittance. 

\item The emittance growth due to synchrotron radiation emission must be constrained within tolerable levels.

\item the half gap $a_x$ must be big enough to minimise the near-axis wakefield effects on the beam during normal operation of the machine.

\end{itemize}

\noindent For a given lattice the horizontal emittance growth due to incoherent synchrotron radiation can be evaluated using the following expression \cite{Sands}:

\begin{equation}
\Delta (\gamma \epsilon_x) \simeq (4.13 \times 10^{-8}~{\textrm m}^2 \textrm{GeV}^{-6}) E^6 I_5 \,\,,
\label{emittancegrowth}
\end{equation}      

\noindent as a function of the beam energy $E$ and the so-called radiation integral $I_5$, which is defined as \cite{Helm},

\begin{equation}
I_5 = \int_{0}^{L} \frac{\cal H}{\left| \rho^{3}_{x} \right|} \,  ds =\sum_{i} L_i \frac{\langle {\cal H} \rangle_i}{\left| \rho^{3}_{x,i} \right|} \; ,
\label{nl:27}
\end{equation}

\noindent where the sum runs over all bending magnets, with bending radius $\rho_i$, length $L_i$, and the average of the function ${\cal H}$, which is defined by:

\begin{equation}
{\cal H}=\frac{D^2_x + (D'_x \beta_x + D_x \alpha_x)^2}{\beta_x}\,\,,
\end{equation} 

\noindent where $\beta_x$ and $\alpha_x=-(1/2)d\beta_x/ds$ denote the typical twiss parameters, $D_x$ the dispersion function and $D'_x =dD_x/ds$.

For the CLIC collimation system $I_5 \simeq 1.9 \times 10^{-19}$~m$^{-1}$, and then $\Delta (\gamma \epsilon_x) \simeq 0.089~\mu$m. This means about $13.5\%$ emittance growth respect to the design emittance $\gamma \epsilon_x = 0.66$~$\mu$m. This corresponds to a beam core luminosity loss of $\Delta \mathcal{L}/\mathcal{L}_0 = 1-1/\sqrt{ 1 +\Delta (\gamma \epsilon_x)/(\gamma \epsilon_{x})} \simeq 6\%$. For the total CLIC BDS (including the CLIC collimation system and the FFS) it results $I_5 \simeq 3.8 \times 10^{-19}$~m$^{-1}$ and an emittance growth of $\Delta (\gamma \epsilon_x)/(\gamma \epsilon_x) \simeq 27.3\%$. This translates into a total luminosity loss of about $11.4\%$. This value is much lower than the result of $24\%$ obtained in Ref.~\cite{DalenaSR} from beam tracking simulations. This discrepancy is basically due to the fact that our calculation from Eqs.~(\ref{emittancegrowth}) only considers the effect from the radiation emission due to the deflection of the beam by the bending magnets, while the tracking simulations also take into account the additional effect from the optical nonlinearities of the lattice.    
     
In the following sections we describe the design of the spoiler and absorber based on survival considerations and, by means of simulations, the thermo-mechanical performance of the spoiler is investigated in detail for the worst damage scenario from a full bunch train impact. Collimation efficiency simulation studies are also performed in order to optimise the collimation apertures.  


\subsection{Spoiler and absorber design}

\begin{figure}[htb]
\begin{center}
\includegraphics*[width=10cm]{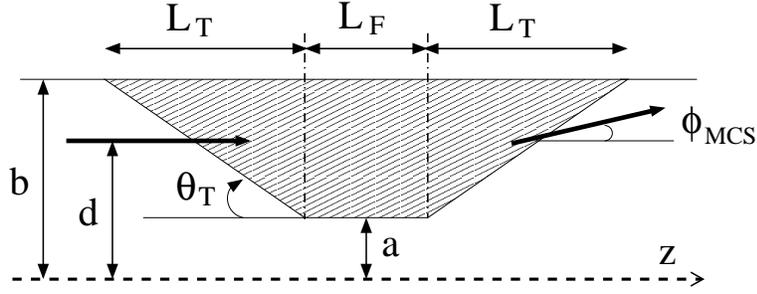}
\caption{Spoiler and absorber jaw longitudinal view.}
\label{spoilergeom}
\end{center}
\end{figure}

This section is devoted to the optimisation of the geometric dimensions of the  energy spoiler and absorber, considering the geometry of Fig.~\ref{spoilergeom}. The design parameters of the energy spoiler and absorber are shown in Table~\ref{tablecoll4}. 

\begin{table}[!htb]
  \caption{Design parameters of the CLIC energy spoiler and absorber.}
  \label{tablecoll4}
  \begin{center}
        \begin{tabular}{l c c}
\hline \hline 
          Parameter & ENGYSP (spoiler) & ENGYAB (absorber) \\      
          \hline 
          Geometry & Rectangular & Rectangular \\
          Hor. half-gap $a_x$ [mm] &  3.51  & 5.41 \\
          Vert. half-gap $a_y$ [mm]  &  8.0 & 8.0\\
          Tapered part radius b [mm] & 8.0 & 8.0 \\
          Tapered part length $L_T$ [mm] & 90.0 & 27.0 \\ 
          Taper angle $\theta_T$ [mrad] & 50.0 & 100.0 \\
          Flat part length $L_F$ [radiation length] & 0.05 & 18.0 \\
          Material & Be & Ti alloy--Cu coating \\                      
    \hline    
    \end{tabular}
  \end{center}
 \end{table}   

\subsubsection{Absorber protection.}
\label{absorberprotection}

The main function of the spoiler is to provide sufficient beam angular divergence by multiple Coulomb scattering (MCS) to decrease the transverse density of an incident beam, thereby reducing the damage probability of the downstream absorber and any other downstream component. This condition determines the minimum length of the material traversed by the beam in the spoiler, i.e. the flat part of the spoiler body ($L_F \neq 0$).

Beam particles traversing the spoiler material are deflected by MCS. The transverse root mean square (rms) scattering angle experienced by the beam particle at the exit of the spoiler can be calculated using the well known Gaussian approximation of the Moli\`ere formula \cite{PDG}:
     
\begin{equation}
\phi_{MCS}=\frac{13.6~[\textrm{MeV}]}{\beta c p} z \sqrt{\frac{\ell}{X_0}} \left[ 1 + 0.038 \ln \left( \frac{\ell}{X_0}\right)\right] \,\, ,
\label{Coulombscatter}
\end{equation}

\noindent where $X_0$ is the radiation length of the spoiler material, $\ell$ is the length of material traversed by the beam particle, $\beta$ is the relativistic factor ($\beta \simeq1$ for ultra-relativistic beams), $c$ the speed of light, $p$ the beam momentum, and $z$ is the charge of the incident particle ($z=1$ for electrons and positrons). Equation~(\ref{Coulombscatter}) is accurate to $11\%$ or better for $10^{-3} < \ell/X_0 < 100$. The square of the transverse angular divergence of a beam at the exit of the spoiler is given by  $\langle x'^2_{sp} \rangle = \langle x'^2_{sp0} \rangle + \phi^2_{MCS}$, and   $\langle y'^2_{sp} \rangle = \langle y'^2_{sp0} \rangle + \phi^2_{MCS}$ for the horizontal and vertical plane, respectively. The terms  $\langle x'^2_{sp0} \rangle$ and $\langle y'^2_{sp0} \rangle$ refer to the initial angular components at the entrance of the spoiler and are usually much smaller than the scattering angular component. Taking into account the linear transport, the expected value of the square of the horizontal and vertical displacements at the downstream absorber can be approximated by 

\begin{eqnarray}
\langle x^2_{ab} \rangle & \simeq & R^2_{12}(s_{sp} \rightarrow s_{ab}) \phi^2_{MCS} + D^2_x \sigma^{2}_{\delta} \,\,, \label{absorbereq1}\\ 
\langle y^2_{ab} \rangle & \simeq & R^2_{34}(s_{sp} \rightarrow s_{ab}) \phi^2_{MCS}\,\,. \label{absorbereq2}
\end{eqnarray} 

In Eq.~(\ref{absorbereq1}) the dispersive component $D^2_x \sigma^2_{\delta}$ has been taken into account, with $D_x =0.416$~m the horizontal dispersion at the energy absorber position, and $\sigma_{\delta}=\sqrt{\langle \delta^2_E \rangle - \langle \delta_E \rangle^2}$ the rms beam energy spread. $\delta_E \equiv \Delta E/E_0$ represents the energy deviation, with $E_0$ the nominal beam energy. $R_{12}(s_{sp} \rightarrow s_{ab})=160.75$~m and $R_{34}(s_{sp} \rightarrow s_{ab})=169.26$~m are the corresponding linear transfer matrix elements between the energy spoiler and absorber. For beam energy 1500~GeV and length of the spoiler material $\ell < 1$~$X_0$ the rms angular divergence by MCS is $\phi_{MCS} \sim 1~\mu$rad. If one considers energy spread values $\sigma_{\delta} \approx 0.29\%$, the energy dispersive term $D_x \sigma_{\delta}$ is dominant in Eq.~(\ref{absorbereq1}), and we can approximate the transverse beam size at the absorber position $s_{ab}$ by:    

\begin{eqnarray}
\sigma_x (s_{ab}) = \sqrt{\langle x^2_{ab} \rangle} & \simeq & D_x \sigma_{\delta} \,\,, \label{absorbereq3}\\ 
\sigma_y (s_{ab}) = \sqrt{\langle y^2_{ab} \rangle} & \simeq & R_{34}(s_{sp} \rightarrow s_{ab}) \phi_{MCS}\,\,. \label{absorbereq4}
\end{eqnarray}     

For the protection of an absorber made of Ti alloy, the following limit for the radial beam size can be established \cite{PT2, PT1}:

\begin{equation}
\sigma_r(s_{ab}) =\sqrt{\sigma_x(s_{ab}) \sigma_y(s_{ab})} \gtrsim 600~\mu\textrm{m} \,\,.
\label{absorbersurvivallimit1}
\end{equation}   

Using Eqs.~(\ref{absorbereq3}) and (\ref{absorbereq4}), the constraint (\ref{absorbersurvivallimit1}) can be rewritten as follows:

\begin{equation}
\sqrt{|R_{34}(s_{sp} \rightarrow s_{ab})| D_x \sigma_{\delta} \phi_{MCS}} \gtrsim 600~\mu\textrm{m} \,\,.
\label{absorbersurvivallimit2}
\end{equation} 

In terms of the transverse particle density peak, the condition for absorber survival can be written as:  

\begin{equation}
\hat{\rho}(s_{ab}) = \frac{N_e}{2\pi \sigma^2_r (s_{ab})} \lesssim 1.64 \times 10^9~\textrm{particles}/\textrm{mm}^2~\textrm{per bunch} \,\,, 
\label{absorbersurvivallimit3}
\end{equation}

\noindent where $N_e=3.72\times 10^9$ is the number of particles per bunch.
 
Considering a Gaussian beam energy distribution with $\sigma_{\delta} = 0.5\%$ energy spread width, from the constraint (\ref{absorbersurvivallimit2}) one obtains that $\phi_{MCS} \gtrsim 10^{-6}$~rad ensures the absorber survival. From this condition and using Eq.~(\ref{Coulombscatter}) one can determine the minimum length of spoiler material necessary to guarantee the absorber survival. This condition is fulfilled if the Be spoiler (Fig.~\ref{spoilergeom}) is designed with a central flat part of length $L_F \gtrsim 0.02~X_0$. Similar results are obtained considering a beam with a uniform energy distribution of $A_{\delta}=1\%$ full energy spread and where $\sigma_{\delta} = A_{\delta}/\sqrt{12}$.

In order to validate these results tracking simulations of bunches have been performed through the CLIC BDS, with 50000 macroparticles per bunch, using the code PLACET \cite{placetoctave}. In this beam model a macroparticle represents a large number of electrons (or positrons) with nearly the same energy and phase space position. For instance, macroparticle $i$ is represented by a 6-D phase space vector $(x_i,x'_i,y_i,y'_i,z_i,E_i)$, by a number of second moments\footnote{The second moments are the covariances of transverse phase coordinates for all particles represented by the macroparticle.}, and by a weight proportional to the number of particles it represents. 

\begin{figure}[htb]
\begin{center}
  \includegraphics*[width=11cm]{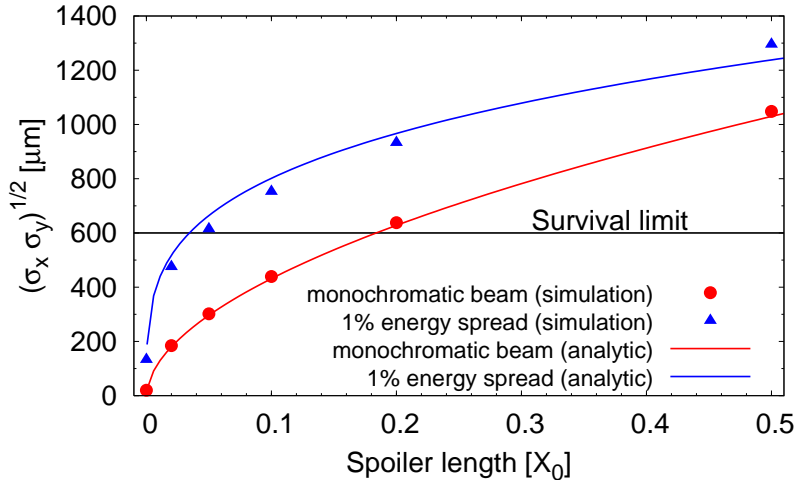}
\caption{Transverse spot size at the energy absorber position as a function of the upstream spoiler length.}
\label{absorberspotsize}
\end{center}
\end{figure}

\begin{figure}[htb]
\begin{center}
  \includegraphics*[width=12cm]{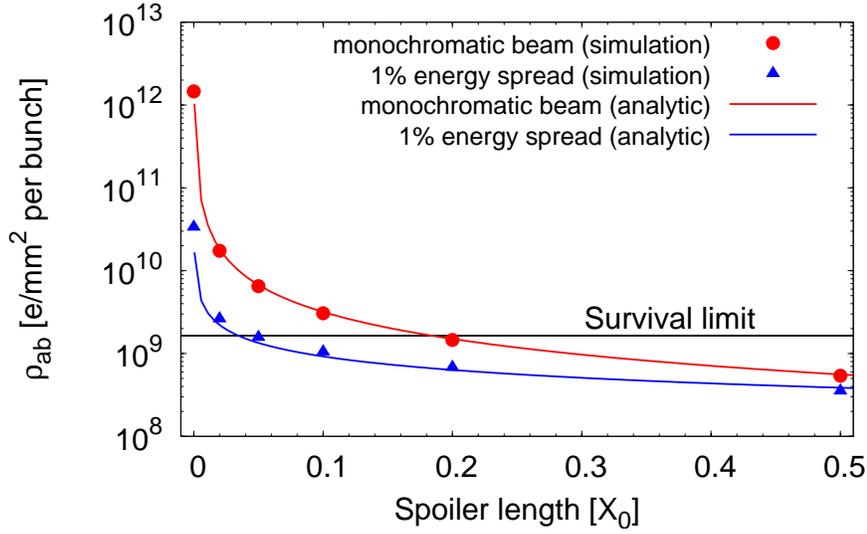}
\caption{Transverse beam density at the energy absorber position as a function of the upstream spoiler length.}
\label{absorberbeamdensity}
\end{center}
\end{figure}

Assuming all particles of the beam hit the energy spoiler and full beam transmission through the spoiler, and applying MCS, we have calculated the transverse beam spot size $\sigma_r=\sqrt{\sigma_x \sigma_y}$ and its corresponding transverse beam density at the energy absorber. From the tracking simulations of the 50000 macroparticles of one bunch, $\sigma_x$ and $\sigma_y$ are calculated from the rms of the $x$ and $y$ positions of the macroparticle distribution. Figure~\ref{absorberspotsize} compares the result of $\sigma_r$ at the absorber position as a function of the spoiler length (in units of radiation length) traversed by the beam for the following cases: a monochromatic beam, i.e. with no energy spread, and a beam with a uniform energy distribution of $1\%$ full spread. The results from the tracking simulations are compared with those from analytical calculations using Eqs.~(\ref{absorbereq3}) and (\ref{absorbereq4}). The corresponding results in terms of transverse particle density are shown in Fig.~\ref{absorberbeamdensity}. For a realistic case of a beam with $1\%$ of energy spread, selecting a length for the energy spoiler of about $0.05~X_0$ might be enough to ensure the survivability of the downstream absorber in case of a full impact of the beam.             

\subsubsection{Spoiler protection.}
\label{spoilersurvival}

Based on the SLC experience\footnote{The Stanford Linear Collider (SLC) \cite{SLC} is the sole linear collider built to date.}, energy errors in the linac are expected to occur much more frequently than orbit disruptions of on-energy beams. Therefore, the E-spoiler has to be designed robust enough so that it survives without damage from the impact of an entire bunch train in case of likely events generating energy errors. 

The instantaneous heat deposition is the principal mechanism leading to spoiler/collimator damage. The main sources of such a heating are the energy deposition by direct beam-spoiler material interaction, the image current heat deposition and the electric field breakdown. The most critical case is the instantaneous temperature rise in the spoiler due to a deep beam impact. Since the thickness of the spoiler is significantly small in terms of radiation length ($L_F \ll 1~X_0$), electrons/positrons deposit energy basically by ionization, and practically no electromagnetic showers are developed. 

As an approximate criterion for spoiler survival the following condition can be established: the instantaneous temperature increment due to the impact of a full bunch train on the spoiler ($\Delta \hat{T}_{\rm inst}$) must be lower than the temperature excursion limit for melting ($\Delta T_{\rm melt}$) and the temperature excursion limit for fracture of the material by thermal stress ($\Delta T_{\rm fr}$), i.e. 

\begin{equation}
\Delta \hat{T}_{\rm inst} = \frac{1}{\varrho C_p} \left(\frac{dE}{dz}\right) \frac{N_e N_b}{2\pi \sigma_x \sigma_y} < \textrm{min} [\Delta T_{\rm fr}, \Delta T_{\rm melt} ] \,\,,
\label{limittemperature}
\end{equation}

\noindent where $\varrho$ is the density of the spoiler material, $C_p$ is the heat capacity, $N_e$ the bunch population and $N_b$ the number of bunches per train.  The safe limit is below the minimum between the thermal stress temperature limit $\Delta T_{\rm fr}$ and the melting limit $\Delta T_{\rm melt}$. Generally the minimum corresponds to $\Delta T_{\rm fr}$. 

Here the energy deposition per unit length is denoted as $(dE/dz)$, whose value can be determined using the formula for the collision stopping power given in Ref.~\cite{Seltzer} in the high energy limit: 

\begin{equation}
\frac{1}{\varrho}\left(\frac{dE}{dz}\right)=0.153536 \frac{Z}{A} B(T)\,\,,
\label{stoppingpower}
\end{equation}

\noindent where $Z/A$ is the ratio of the number of electrons in the atom to the atomic weight of the spoiler material, and $B(T)$ is the stopping number defined in \cite{Seltzer}. It is necessary to mention that Eq.~(\ref{stoppingpower}) gives a conservative estimation of the energy deposited in the spoiler and overestimates it, since by definition the stopping power is the energy lost by the passing beam, and not the energy that is actually deposited in the target. A fraction of the lost energy might indeed escape from the spoiler.

Table~\ref{tabletemperature1} shows the instantaneous increment of temperature calculated using Eqs.~(\ref{limittemperature}) and (\ref{stoppingpower}) for CLIC electron and positron beams and for different spoiler materials. For these calculations we have neglected the temperature dependence of the heat capacity $C_p$ and used the following rms transverse beam sizes: $\sigma_x=779.6~\mu$m and $\sigma_y=21.9~\mu$m. The material properties of Table~\ref{materialproperties} have been considered. These material data have been obtained from Ref.~\cite{materialweb}.      

\begin{table}[!htb]
\caption{Energy deposition per unit length ($dE/dz$) estimated from Eq.~(\ref{stoppingpower}) for a CLIC beam traversing a thin spoiler, and instantaneous temperature increment calculated using Eq.~(\ref{limittemperature}). Different spoiler materials are compared.}
\label{tabletemperature1}
\begin{center}
\begin{tabular}{|c|cc|cc|}
\hline \hline 
Spoiler  & \multicolumn{2}{c}{\bf Electron beam} & \multicolumn{2}{|c|}{\bf Positron beam} \\
Material & $dE/dz$ [MeV/cm] & $\Delta \hat{T}_{inst}$ [K] & $dE/dz$ [MeV/cm] & $\Delta \hat{T}_{inst}$ [K] \\
\hline
Be & 4.4003 & 214 & 4.3181 & 209 \\
C & 6.001 & 648 & 5.8879 & 636 \\
Ti & 10.8487 & 786 & 10.6406 & 770 \\
Cu & 20.9522 & 1049 & 20.5422 & 1028 \\
W & 39.1714 & 2606 & 38.3897 & 2554 \\
\hline 
\end{tabular}
\end{center}
\end{table}

\begin{table}[!htb]
\caption{Material properties: atomic number $Z$, mass number $A$, material density $\varrho$, specific heat capacity $C_p$, electrical conductivity $\sigma$ (at room temperature, 293~K) and radiation length $X_0$.}
\label{materialproperties}
\begin{center}
\begin{tabular}{lccccccc}
\hline \hline 
Material & $Z$ & $A$ [g/mol] & $\varrho$ [gm$^{-3}$] & $C_p$ [Jg$^{-1}$K$^{-1}$] & $\sigma$ [$\Omega^{-1}$m$^{-1}$] & $X_0$ [m] \\
\hline
Be & 4 & $9.01218$ & $1.84 \times 10^6$ & $1.925$ & $2.3 \times 10^7$ & $0.353$ \\
C &  6 & $12.0107$ & $2.25\times 10^6$ & $0.708$  & $1.7 \times 10^4$ & $0.188$ \\
Ti & 22 & $47.867$ & $4.5 \times 10^6$ & $0.528$ & $1.8 \times 10^6$ & $0.036$ \\
Cu & 29 & $63.546$ & $8.93 \times 10^6$ & $0.385$ & $5.9 \times 10^7$ & $0.014$ \\
W & 74 & $183.84$ & $19.3 \times 10^6$ & $0.134$ & $1.8 \times 10^7$ & $0.0035$ \\
\hline 
\end{tabular}
\end{center}
\end{table}

The rapid heating of the material caused by the impact of the train in the spoiler may contribute to the fracture of the material by thermal stress. The increment of temperature which determines the limit for thermal fracture can be analytically evaluated using the following expression:

\begin{equation}
\Delta T_{\rm fr} \cong \frac{2 \sigma_{\rm UTS}}{\alpha_T Y}\,\,,
\label{fracturelimit}
\end{equation}

\noindent where $\sigma_{\rm UTS}$ is the ultimate tensile strength, $\alpha_T$ is the thermal expansion coefficient and $Y$ is the modulus of elasticity (or Young modulus). The ultimate tensile strength is defined as the maximum stress that the material can withstand. It is necessary to mention that for the value of $\sigma_{\rm UTS}$ discrepancies of up to $40\%$ can be found between different bibliographic sources about material data. Here we have used the material information from Ref.~\cite{materialweb}, which gives a pesimitic value for $\sigma_{\rm UTS}$ in comparison with other bibliographic sources.   

For the CLIC energy spoiler made of Be, using the mechanical and thermal properties of Table~\ref{Beproperties}, we obtain $\Delta T_{\rm fr} \simeq 228$~K, which is slightly bigger than the values obtained for $\Delta \hat{T}_{inst}$ for a Be spoiler (see Table~\ref{tabletemperature1}). Therefore, according this analytic calculation the Be spoiler is below, but close, the fracture limit in case of the impact of an entire CLIC bunch train. 

\begin{table}[!htb]
\caption{Summary of material properties for beryllium.}
\label{Beproperties}
\begin{center}
\begin{tabular}{lc}
\hline \hline 
Young modulus, $Y$ [$10^5$~MPa] & 2.87 \\
Thermal expansion coefficient, $\alpha_T$ [$10^{-6}$~K$^{-1}$] & 11.3 \\
Ultimate tensile strength, $\sigma_{\rm UTS}$ [MPa] & 370 \\
Tensile yield strength [MPa] & 240 \\
Compressive yield strength [MPa] & 270 \\
Specific heat capacity, $C_p$ [J$/$(gK)] & 1.925 \\
Density, $\varrho$ [g/cm$^3$] & 1.84 \\
\hline 
\end{tabular}
\end{center}
\end{table}
 
In general Eq.~(\ref{fracturelimit}) may be a good approximation to estimate the temperature at which the material may crack. However, it is necessary to point out that Eq.~(\ref{fracturelimit}) is commonly used with quasi-static material data and for fatigue purposes. In the case of the spoiler heating by the beam we are not involved in a fatigue process but in a ``one-time'' accident scenario. It is known that when a beam hits a material the energy is deposited very quickly into it. This causes a rapid expansion of the material, and hence quasi-static material properties will not give an accurate answer. In this case, the materials under study need to be characterised dynamically in order to give more valid results. For a more precise thermo-mechanical characterisation of the spoiler material numerical simulations are usually performed using tools such as FLUKA \cite{FLUKA} and ANSYS \cite{ANSYS}. Simulation results are shown in the next section.  

\subsubsection{Thermo-mechanical analysis of the spoiler.}
\label{thermo-mechanical}

In order to evaluate the robustness of the spoiler, simulations, using the codes FLUKA and ANSYS, have been made considering the geometrical parameters of the CLIC E-spoiler made of Be, and assuming the nominal parameters of the CLIC beam. 

\begin{figure}[htb]
\begin{center}
  \includegraphics*[width=12cm]{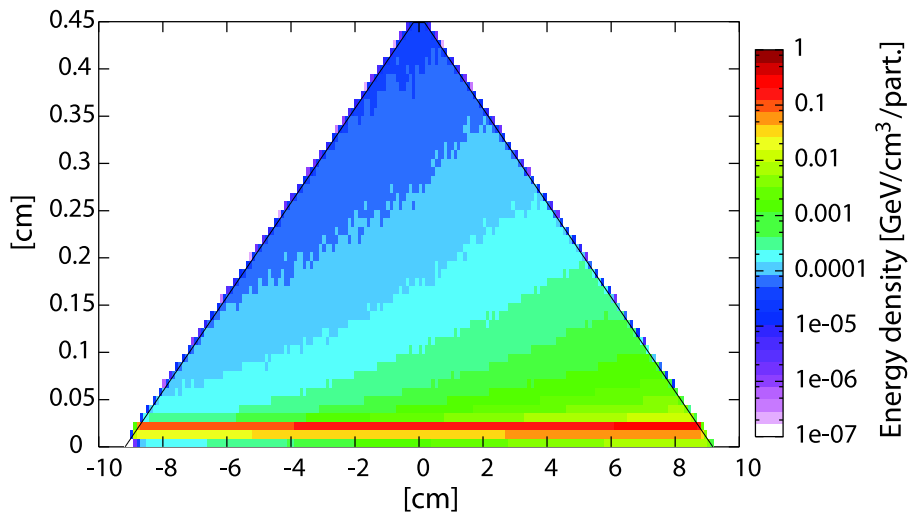}
  \caption{Energy density deposition normalised per incident particle for a CLIC beam hitting the spoiler.}
\label{energydeposition1}
\end{center}
\end{figure}

\begin{figure}[htb]
\begin{center}
  \includegraphics*[width=12cm]{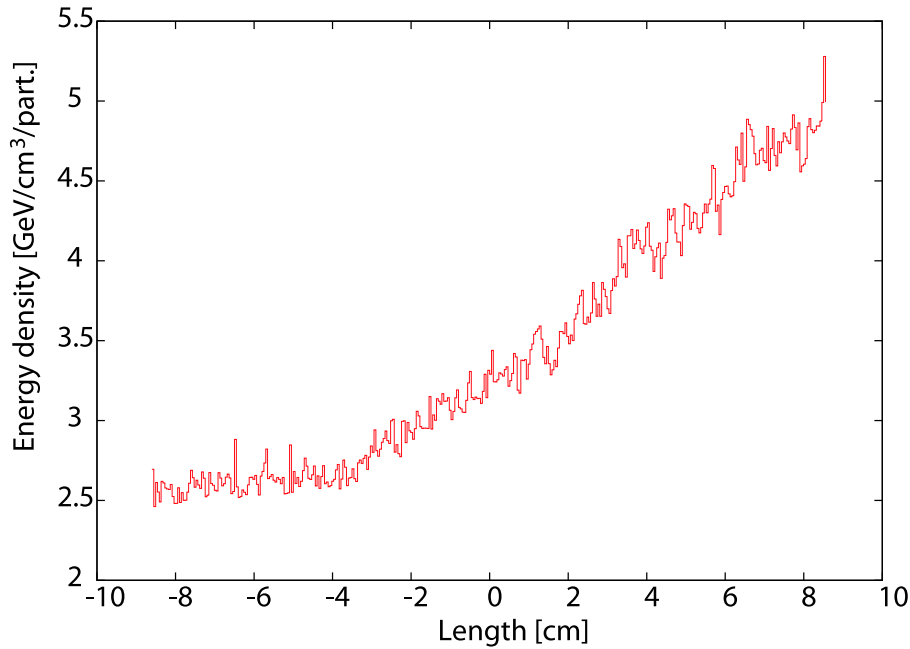}
  \caption{Peaks of energy density deposition normalised per incident particle for a CLIC beam hitting the spoiler.}
\label{energydeposition2}
\end{center}
\end{figure}

The following horizontal and vertical beam sizes at the spoiler position have been assumed\footnote{$\sigma_x=779~\mu$m at spoiler position corresponds to the rms horizontal beam size of a beam with a uniform energy spread of $1\%$ full width. However, in this FLUKA simulation we have assumed the nominal energy for all particles of the beam and no energy spread. This assumption gives more pessimistic predictions than a more realistic situation.}: $\sigma_x=779~\mu$m and $\sigma_y=21.9~\mu$m. The bunch train impact was simulated using FLUKA. Figure~\ref{energydeposition1} shows the energy deposition in the spoiler as the beam traverses it. A transverse position depth of $d \simeq 7.8$~mm (see Fig.~\ref{spoilergeom}) for the beam was chosen, to maximise the total amount of material that it would face in case of a pessimistic accident scenario. This represents a deviation of about $10~\sigma_x$ from nominal orbit. Figure~\ref{energydeposition2} shows the corresponding peaks of energy density along the beam track in the spoiler material. The peak of energy deposition happens in the edge of the trailing taper and is about $5.4$~GeV$/$cm$^3$ per incident particle; using the specific heat and density values of beryllium, shown in Table~\ref{Beproperties}, and the total number of particles in a CLIC bunch train, $N_b \times N_e=1.16 \times 10^{12}$, a temperature increment of approximately 570~K is obtained. 

\begin{figure}[htb]
\begin{center}
\includegraphics*[width=8cm]{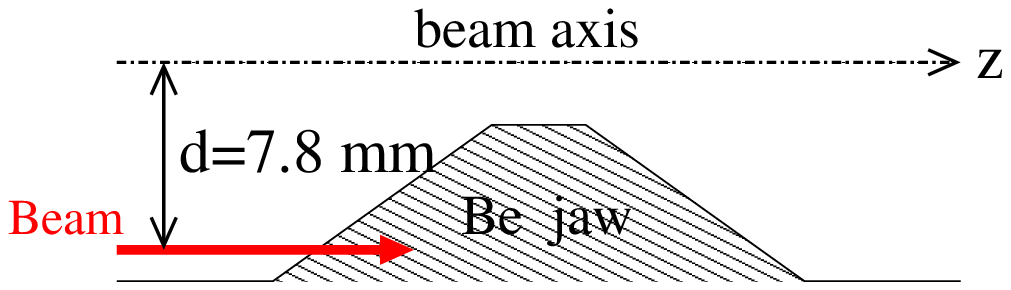}
\includegraphics*[width=12cm]{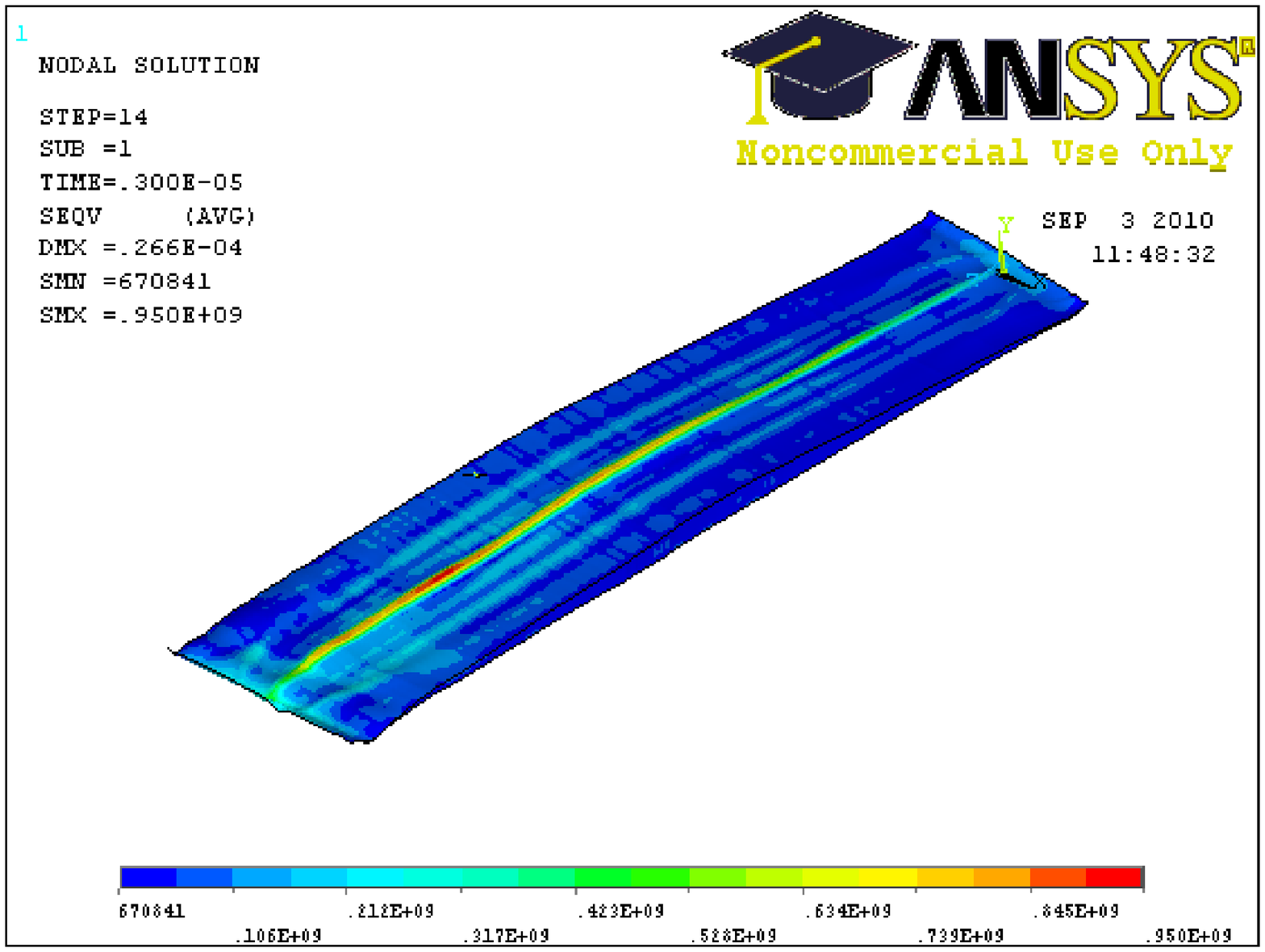}
  \caption{Equivalent stress on the spoiler body 3 microseconds after a CLIC bunch train hits it. In this case the transverse impact depth is $d=7.8$~mm, which corresponds to a beam deviation of $10$~$\sigma_x$ with respect to the beam axis.}
\label{ANSYSplot1}
\end{center}
\end{figure}

In order to perform the transient analysis of the CLIC train hitting the Be E-spoiler, the FLUKA result was transformed into an ANSYS input and applied in a spoiler model. The results are recorded after the beam has hit the spoiler to determine if there would be any stress build up that could reach fracture levels. The results of the stress calculations in the Be can be compared with the mechanical stress limits of the material by means of a certain failure criterion expressed by the equivalent stress values\footnote{This equivalent stress is also called \emph{von Mises stress} \cite{Mises}, and is often used for metals under multi-axial state stress. It allows any arbitrary three-dimensional stress state to be represented as a single positive stress value. Equivalent stress is part of the maximum equivalent stress failure theory used to predict the onset of yielding and to describe the post-yielding response.}:

\begin{equation}
\sigma_{eq}=\frac{1}{\sqrt{2}}\sqrt{(\sigma_1-\sigma_2)^2 + (\sigma_2-\sigma_3)^2 + (\sigma_3-\sigma_1)^2}\,\,,
\label{equivalentstress}
\end{equation}   

\noindent where $\sigma_1$, $\sigma_2$ and $\sigma_3$ are the principal stresses at a given position in the three main directions of the working coordinate system, which in our case is Cartesian. Figure~\ref{ANSYSplot1} shows the equivalent stress calculated using ANSYS on the spoiler body 3~$\mu$s after the full CLIC bunch train has hit it, time at which the stress reaches its maximum and stabilises, with an impact depth of $d \simeq 7.8$~mm. In this case we obtain a top equivalent stress of $\approx 950$~MPa, and tensile, which is way above the ultimate tensile strength limit, thus reaching fracture levels.      

\begin{figure}[htb]
\begin{center}
 \includegraphics*[width=8cm]{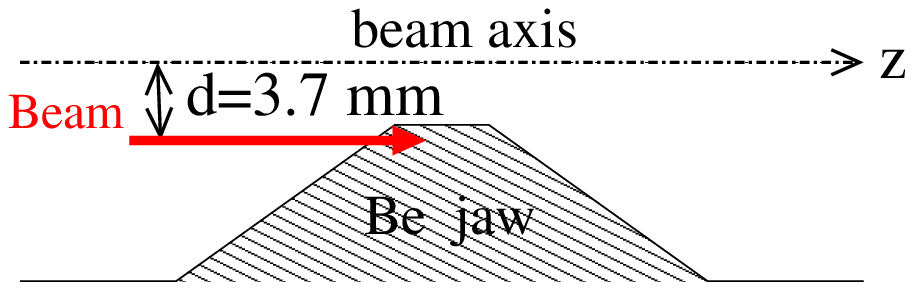}
 \includegraphics*[width=12cm]{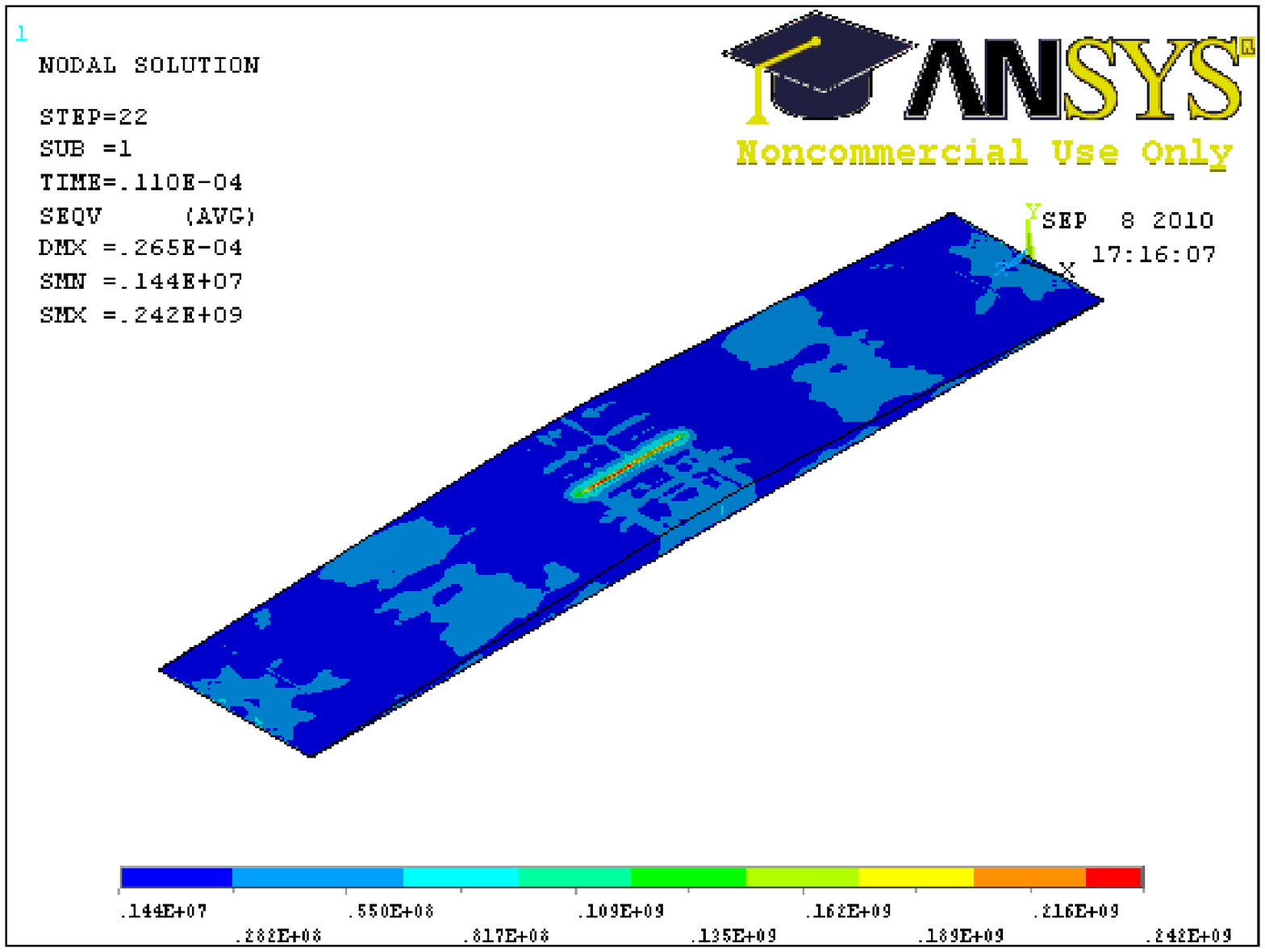}
  \caption{Equivalent stress on the spoiler body 11 microseconds after a CLIC bunch train hits it. In this case, the transverse impact depth is $d=3.7$~mm, which corresponds to a beam deviation of $4.8$~$\sigma_x$ with respect to the beam axis.}
\label{ANSYSplot2}
\end{center}
\end{figure}

Let us now consider another case of impact in which the beam traverses less quantity of material. For instance, the case of the beam hitting the spoiler with impact depth $d=3.7$~mm, which means a deviation of about $5~\sigma_x$ with respect to the nominal beam axis. Figure~\ref{ANSYSplot2} shows the equivalent stress  calculated using ANSYS, after 11~$\mu$s, the time needed in this case for the stress to reach its peak and stabilise over that top value. The maximum value of stress after a CLIC bunch train has hit the spoiler is about 240~MPa, and compressive, value that corresponds to the yield compressive strength value. In this situation there will not be fracture, but there might be a permanent deformation. This deformation translates into horizontal protuberances of $\sim 1~\mu$m, which represents $0.03\%$ of the minimum half gap of the E-spoiler. This might have consequences in terms of degradation of the beam stability and emittance blow-up by increasing the collimator wakefield effects. The additional wakefield effects due to the deformation of the spoiler are evaluated in Section~3.3 (Part~II).
     
Above we have considered two cases of impact position on the spoiler surface: a big transverse impact depth of about $10~\sigma_x$ from the nominal beam axis, and a more optimistic scenario with an impact depth of about $5~\sigma_x$. These  two examples, one more pessimistic than the other, have allowed us to obtain a preliminary estimate of the survivability of the CLIC E-spoiler. However, the impact position of the beam on the spoiler surface depends on failure scenarios, and a detailed study of these failure events affecting the beam energy would be useful in order to determine the most likely angles and positions of impact for a more precise risk analysis. 

Another necessary remark is that this study has been performed for a perfect beryllium structure, i.e. without any imperfections or impurities, which could act as a stress concentrator. Therefore, Be samples will need to be tested to compressive stress up to 200~MPa to assess their suitability for spoiler manufacturing.          

\subsection{Collimation efficiency}

In this section the capability of the system to intercept off-energy beams is investigated by means of particle tracking simulations. 

\begin{figure}[htb]
\begin{center}
  \includegraphics*[width=7cm]{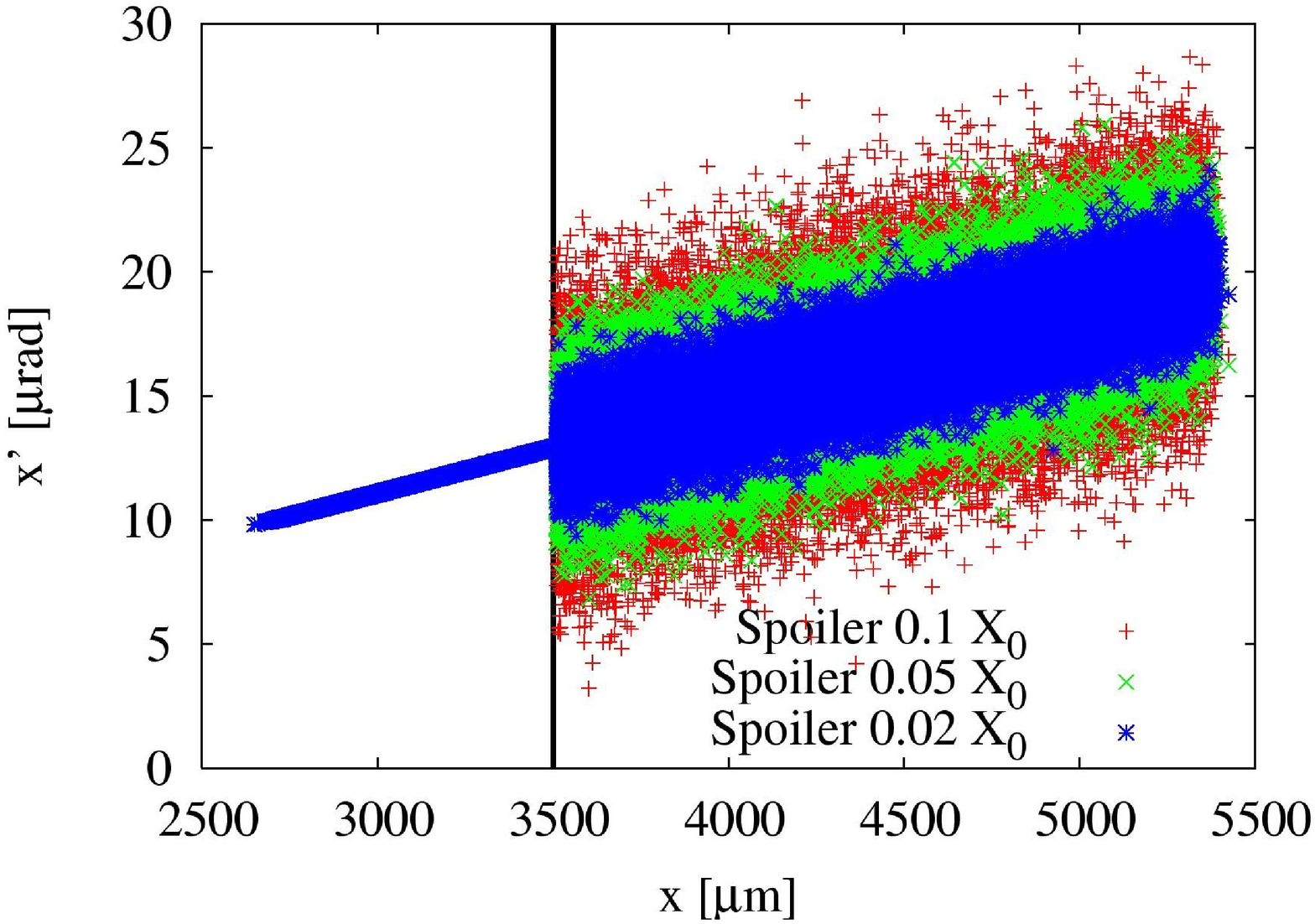}
  \includegraphics*[width=7cm]{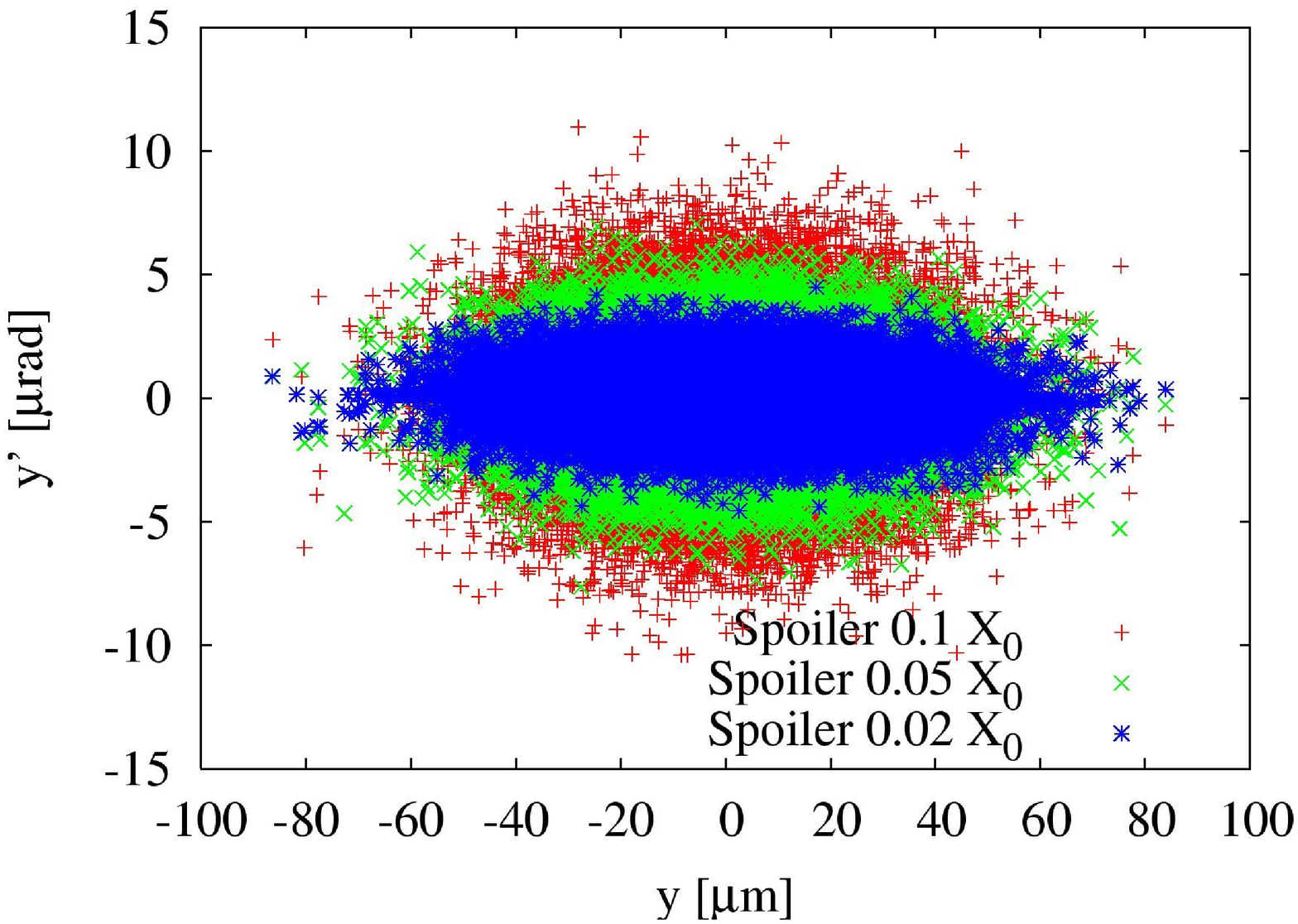} 
\caption{Left: $x$--$x'$ phase space at the exit of the spoiler. Right: $y$--$y'$ phase space at the exit of the spoiler. The following cases of traversed spoiler length are represented: $0.02~X_0$, $0.05~X_0$ and $0.1~X_0$. The collimation limit, determined by the edge of the spoiler jaw, is represented by the vertical black line.}
\label{spoilerexit}
\end{center}
\end{figure}

Let us assume complete transmission of the beam through the E-spoiler\footnote{This approximation is only valid for very thin spoilers (less than 1 radiation length) made of materials with low $Z$.} and perfect collimation at the absorber, i.e. particles hitting the absorber are considered totally lost without production of secondary particles. With these assumptions, a beam of initial energy offset $1.5\%$ of the nominal energy and $1\%$ full energy spread has been tracked through the CLIC BDS using the code PLACET. Figure~\ref{spoilerexit} shows the horizontal and vertical phase space at the exit of the E-spoiler, taking into account the effect of MCS for different cases of traversed spoiler length in units of radiation length ($X_0$). The tracking results show how the transverse beam phase space area increases at the spoiler exit as the spoiler length increases. In Fig.~\ref{spoilerexit} (Left) the results also show that part of the beam (with $x$ amplitude $< 3.5$~mm) does not hit the spoiler and is not scattered by MCS. To avoid this, if we demand a complete interception of the off-energy beam (with the above energy conditions), the E-spoiler half gap has to be reduced further, to about 2.5~mm. Reducing the spoiler half gap, the wakefield effects increase. This may be a possible cause for concern. However, as we will see in Section~3 (Part~II), the contribution of the E-spoiler to the wakefields is practically negligible due to its relative large half gap (3.5~mm) in comparison with that of the betatron spoilers ($\sim 100~\mu$m), which significantly contribute to the collimator wakefields for small position offsets from the orbit axis. Reducing the E-spoiler half gap to 2.5~mm might still give a tolerable stability margin in terms of wakefields.     
       
\begin{figure}[htb]
\begin{center}
  \includegraphics*[width=7.6cm]{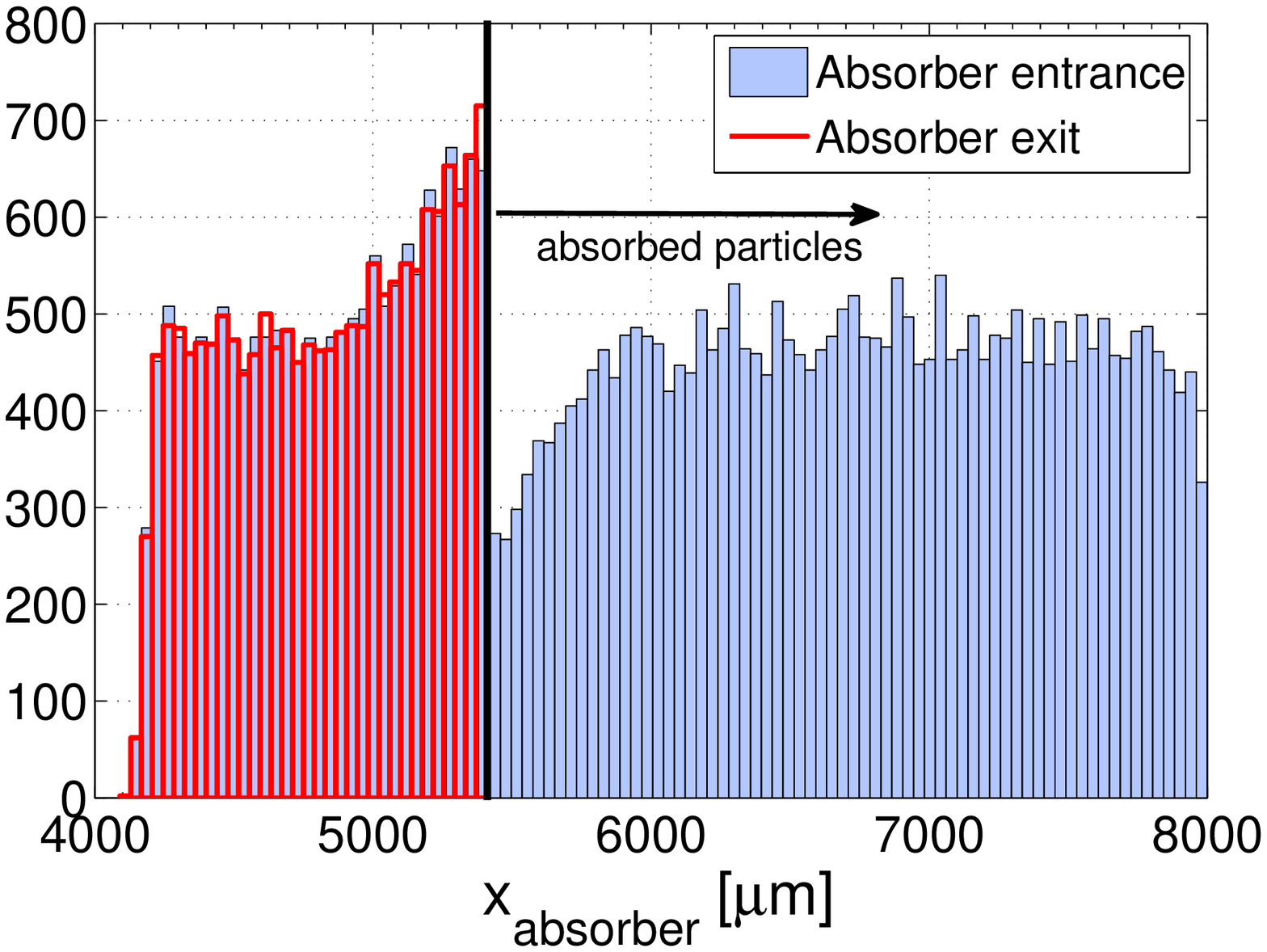}
  \includegraphics*[width=7.6cm]{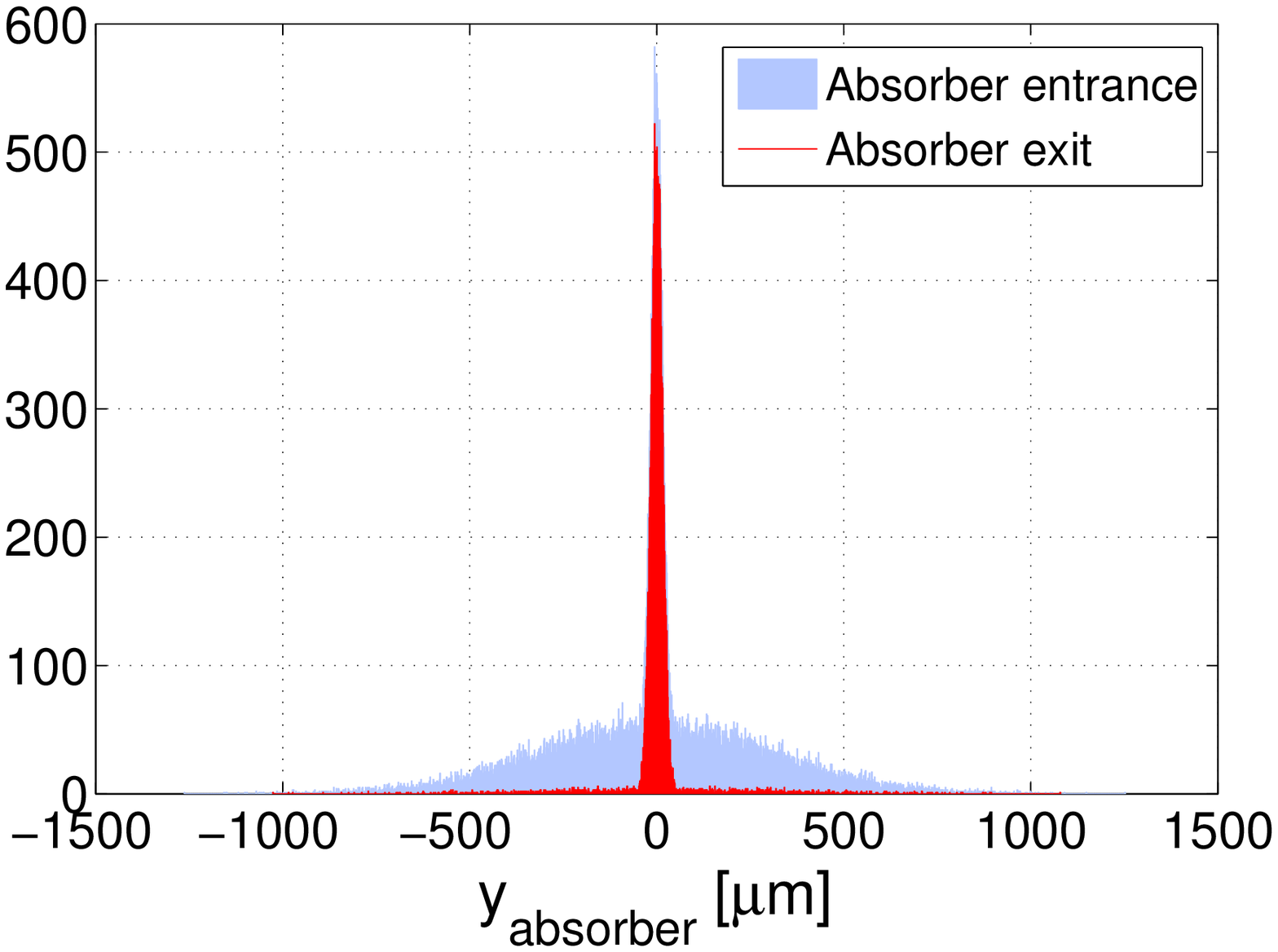} 
\caption{Transverse beam distribution at the E-absorber entrance and exit, considering a beam with $1.5\%$ centroid energy offset and a uniform energy distribution with $1\%$ full width of energy spread. Projection on the horizontal plane (Left), and projection on the vertical plane (Right). The collimation limit determined by the edge of the absorber jaw is represented by the vertical black line.}
\label{absorberprojection_off}
\end{center}
\end{figure}

\begin{figure}[htb]
\begin{center}
  \includegraphics*[width=9cm,angle=90]{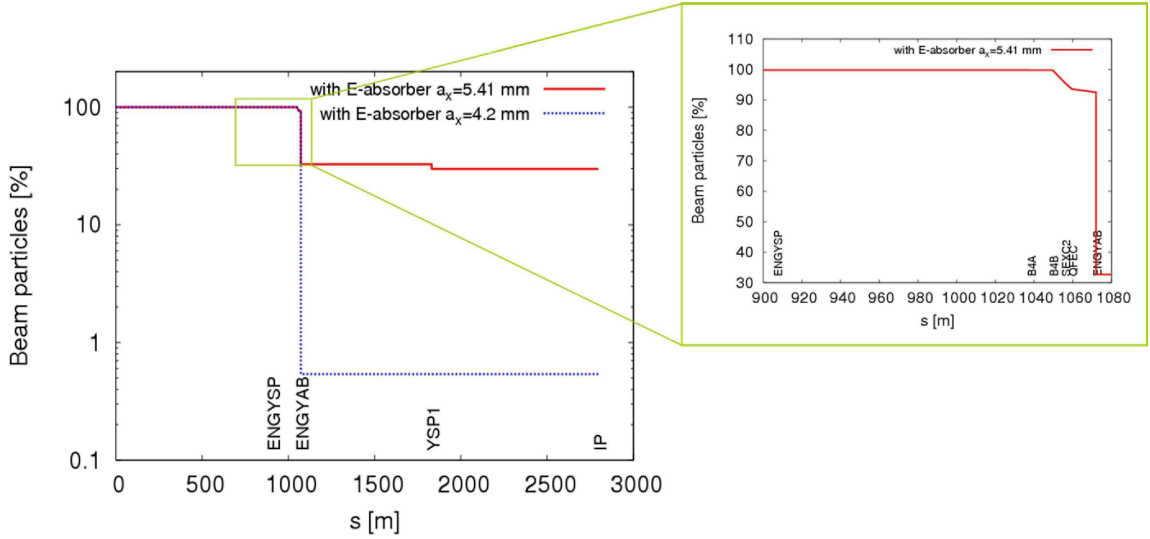}
\vspace{-1.5cm}
 \caption{Left: number of beam particles along the CLIC BDS, considering an initial beam composed by 50000 macroparticles with $1.5\%$ centroid energy offset and $1\%$ full width of energy spread. Multiple Coulomb scattering within the E-spoiler (ENGYSP) increases the angular divergence. Perfect absorption of the beam is considered at the downstream E-absorber (ENGYAB) and at other limiting apertures of the lattice. Notice the logarithmic vertical scale. The cases for E-absorber apertures $a_x=5.41$~mm and $a_x=4.2$~mm are shown. Right: zoom of the particle losses in the section between ENGYSP and ENGYAB.} 
\label{absorber_collimation_efficiency}
\end{center}
\end{figure}

Figure~\ref{absorberprojection_off} shows the horizontal and vertical distribution of the beam particles at the E-absorber. Particles with amplitude $x > 5.41$~mm are perfectly absorbed. However, part of the beam does not hit the absorber jaw and is propagated downstream, with risk of hitting some sensible components of the lattice or at the interaction region. Where are these particles deposited? In order to study the efficiency of the energy collimation system to intercept a miss-steered beam with centroid energy offset $\gtrsim 1.5\%$, the particle loss map along the CLIC BDS has been studied via tracking simulations. As expected, the main particle losses are concentrated at the absorber (see Fig.~\ref{absorber_collimation_efficiency}). However, with the current absorber aperture, $a_x=5.41$~mm, only $70\%$ of the miss-steered beam is collimated. Considering a beam pipe radius of 8~mm in the BDS, approximately $10\%$ of beam losses occur in a region just upstream of the E-absorber. These residual losses of primary beam particles in non-dedicated collimation places (uncontrolled losses) hit the beam-pipe or other parts of lattice elements, thus creating additional fluxes of muons and other secondary particles which propagate downstream. To avoid uncontrolled particle losses, a possible solution could be the increase of the beam pipe radius from the current design 8~mm to a new value of about 10~mm \footnote{Parallel and complementary studies, based on resistive wall effect in the CLIC BDS, have also suggested an optimum beam pipe radius of 10~mm \cite{Mutzner}.}.
    
If the absorber aperture is reduced to $a_x=4.0$~mm, practically $100\%$ of the beam is stopped at the E-absorber.         

\section{Beam diagnostics in the collimation section}

It is planned to set up beam position monitors (BPMs) at every quadrupole of the CLIC BDS and, therefore, each quadrupole of the collimation system will be equipped with one BPM of about 20--50 nm resolution. Sub-micron resolution levels can be achieved by using cavity BPMs. C-band and S-band type BPMs have been successfully commissioned and tested at the KEK final focus Accelerator Test Facility (ATF2) \cite{ATF2}, achieving resolutions in the range 20-200 nm \cite{Boogert}. These BPMs will play a key role in the beam based alignment procedure of the collimation system and, in general, of the whole BDS. They will further form part of the necessary equipment for the implementation of orbit feedback systems for the BDS.

Beam loss monitors (BLMs) distributed along the collimation system would be useful to quantify the beam losses. These BLMs would be integrated into a global machine protection system, which would abort machine operation or activate the necessary protection mechanisms if intolerable levels of radiation are detected. The details of this system will be specified during the technical design phase.                      

\begin{figure}[htb]
\begin{center}
\includegraphics*[width=14cm]{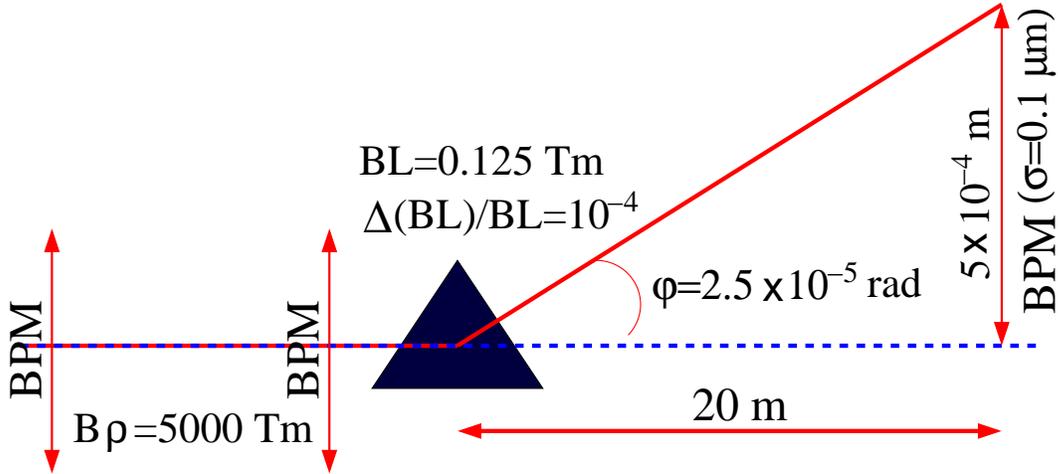}
\caption{Conceptual compact CLIC energy measurement.}
\label{energyspectrometer}
\end{center}
\end{figure}

Another important diagnostic instrument foreseen to be located into the collimation section is the post-linac energy spectrometer. The post-linac energy measurement has been devised in a way to minimise the required space due to the tight constraints in the CLIC total length. The deflection by the first dipole in the energy collimation section, together with three high precision BPMs, provides a compact spectrometer for energy measurement. A conceptual layout of this system is shown in Fig.~\ref{energyspectrometer}. The energy measurement resolution of the set up is estimated to be $\approx 0.04\%$. The integrated magnetic field is assumed to have a calibration error of $\Delta (B L)/BL \approx 0.01\%$ and the BPM resolution is 100~nm. In addition, the energy collimation lattice incorporates a pulsed kicker magnet and a beam dump point, which can extract the beam dowstream of the energy diagnostic station. This permits the linac commissioning without requiring the beam to pass through the IP. 

The CLIC energy collimation section has also a suitable drift space between the collimation dipoles to locate an upstream Compton based polarimeter \cite{Schuler}. It consists of a laser crossing point at position $s=742$~m and a Compton electron detector at $s=907$~m, behind 12 dipoles. This system would allow polarimetry from $1.5$ TeV down to 135 GeV beam energies, but would require several wide-aperture dipoles. If we decide to locate the detector behind a lesser number of dipoles, the dipole aperture requirement would be reduced at the expense of reducing the reachable energy range, e.g. from 1.5 TeV down to 511 GeV, if the detector is placed behind 6 dipoles from the laser position. Ref.~\cite{Schuler} concludes that for CLIC a standard Q-switched YAG laser operated with 100 mJ$/$pulse at 50 Hz would give adequate polarimeter performance. 

\section{Summary and conclusions}

The post-linac collimation system of CLIC must fulfil two main functions: the minimisation of the detector background at the IP by the removal of the beam halo, and the protection of the BDS and the interaction region against miss-steered  or errant beams.   

Recently several aspects of the CLIC post-linac collimation system at 3 TeV CM energy have been optimised in order to improve its performance. This report has been devoted to explain the optimisation procedure and to describe the current status of the CLIC collimation system.   

The CLIC collimation system consists of two sections: one for momentum collimation and another one for betatron collimation. Next, the conclusions for the energy or momentum collimation system are summarised:

\begin{itemize}
\item The energy collimation system of CLIC is designed to remove particles with off-energy $\gtrsim 1.3\%$ of the nominal beam energy. Furthermore, it is conceived as a system for passive protection against beams with large energy offsets ($\gtrsim 1.3\%$), caused by likely failure modes in the main linac. 

\item The design and optimisation of the energy collimators (spoiler and absorber) have been based on survival conditions. The energy collimators are required to survive the impact of an entire bunch train.

\item A minimum spoiler length of $0.05~X_0$ seems to provide enough  transverse angular divergence by MCS to reduce the transverse beam density and guarantee the survivability of the downstream absorber in case of the impact of a bunch train.

\item Beryllium has been selected to made the energy spoiler due to its high thermo-mechanical robustness as well as its high electrical conductivity (to reduce resistive wakefields) in comparison with other materials.

\item Thermo-mechanical studies of the energy spoiler, based on the codes FLUKA \cite{FLUKA} and ANSYS \cite{ANSYS}, have shown that fracture levels are reached if a bunch train hits the spoiler at $\sim 10~\sigma_x$ horizontal offset from the beam axis. In the case of a more optimistic risk scenario, when a bunch train hits the spoiler at $\sim 5~\sigma_x$, practically at the edge of the spoiler, the material does not fracture, but there might be permanent deformations. These deformations consist of horizontal protuberances of $\sim 1~\mu$m. In principle, in terms of near-axis wakefields, a rough evaluation of the consequences of these deformations indicate negligible effects. However, for a more precise evaluation, further studies of near-axis and near-wall wakefield effects are needed.

\item From collimation efficiency studies, based on tracking simulations, the following conclusions can be drawn: increasing the beam pipe aperture from 8~mm to 10~mm seems to help to eliminate undesired residual beam losses in non-dedicated collimation places; reducing the energy collimator half gaps to 2.5 mm (spoiler) and 4 mm (absorber) has proved an optimal removal of beams with $1.5\%$ mean energy offset and $1\%$ full energy spread (for a uniform energy distribution).

\end{itemize}

\section*{Acknowledgements}
This work is supported by the European Commission under the FP7 Research Infrastructures project EuCARD, grant agreement no. 227579.

\end{document}